\documentclass[reprint,aps,twocolumn]{revtex4}

\usepackage{graphicx}
\usepackage{dcolumn}
\usepackage{bm}
\usepackage[dvipsnames]{xcolor}

\usepackage{subfigure}
\usepackage{wrapfig}
\usepackage{cancel}
\usepackage{color}
\usepackage{textcomp}
\usepackage{amsmath}
\usepackage{amsfonts}
\usepackage{amssymb}
\usepackage{array}

\newcommand{\rev}[1]{\textcolor{black}{#1}} 
\newcommand{\CS}{CaSb$_2$}

\newcommand{\Cp}{$C_P$}
\newcommand{\Ce}{$C_{\rm{el}}$}
\newcommand{\Tc}{$T_{\text c}$}

\newcommand{\mHct}{$\mu_{\rm{0}} H_{\rm{c2}}(0)$} 
\newcommand{\mHco}{$\mu_{\rm{0}} H_{\rm{c1}}(0)$}
\newcommand{\mHsco}{$\mu_{\rm{0}} H^{\rm{*}}_{\rm{c1}}(0)$}

\begin{document}

\preprint{APS/123-QED}

\title{Superconductivity and Quantum Oscillations \\in Single Crystals of the Compensated Semimetal \CS\ }

\author{M. Oudah$^{1,2,c}$}
 \email{mohamed.oudah@ubc.ca}
 \thanks{$^{c}$These two authors contributed equally}
\author{J. Bannies$^{1,3,c}$}%
\author{D. A. Bonn$^{1,2}$}%
\author{M. C. Aronson$^{1,2}$}%
\affiliation{%
$^1$Stewart Blusson Quantum Matter Institute, University of British Columbia, Vancouver, British Columbia V6T 1Z4, Canada\\
$^2$Department of Physics, University of British Columbia, Vancouver, Canada V6T 1Z1, Canada\\
$^3$Department of Chemistry, University of British Columbia, Vancouver, Canada V6T 1Z1, Canada\\
}%

\date{\today}

\begin{abstract}
Bulk superconductivity in a topological semimetal is a first step towards realizing topological superconductors, which can host Majorana fermions allowing us to achieve quantum computing. Here, we report superconductivity and compensation of electrons and holes in single crystals of the nodal-line semimetal \CS . We characterize the superconducting state and find that Cooper pairs have moderate-weak coupling, and the superconducting transition in specific heat down to 0.22~K deviates from that of a BCS superconductor. The non-saturating magnetoresistance and electron-hole compensation at low temperature are consistent with density functional theory (DFT) calculations showing nodal-line features. Furthermore, we observe de Haas–van Alphen (dHvA) oscillations consistent with a small Fermi surface in the semimetallic state of \CS . Our DFT calculations show that the two electron bands crossing the Fermi level are associated with Sb1 zig-zag chains, while the hole band is associated with Sb2 zig-zag chains. The Sb1 zig-zag chains form a distorted square net, which may relate the $M$Sb$_2$ family to the well known $M$SbTe square net semimetals. Realization of superconductivity and a compensated semimetal state in single crystals of CaSb$_2$ establishes the diantimonide family as a candidate class of materials for achieving topological superconductivity.
\end{abstract}

\pacs{Valid PACS appear here}

\maketitle

\section{\label{intro}Introduction}

A central theme in the field of topological materials has been the search for topological superconductors, which are expected to host Majorana fermions and have potential applications in achieving fault-tolerant quantum computing~\cite{beenakker2013search,alicea2012new}. 
One avenue for finding topological superconductivity relies on making a heterostructure of a superconductor and a topologically non-trivial material~\cite{fu2008superconducting}, in the hopes of inducing superconductivity via the proximity effect. \rev{In parallel, the quest to find bulk topological superconductors continues~\cite{fu2010odd}, and candidate materials include Cu$_x$Bi$_2$Se$_3$~\cite{sasaki2011topological}, PdBi$_2$~\cite{matthias1963th,imai2012superconductivity}, LaPt$_3$P~\cite{takayama2012strong}, and UTe$_2$~\cite{ran2019nearly}.} The search for superconductivity in topologically non-trivial materials includes work on topological insulators~\cite{hasan2010colloquium} and topological semimetals~\cite{yan2017topological}. In topological insulators, superconductivity can be achieved by doping the material, but this approach has the disadvantage that doping can push the topological states away from the Fermi level~\cite{kondo2013anomalous}. 

In topological semimetals on the other hand, superconductivity may be realized without doping due to the finite density of states at the Fermi level $E_{\rm{F}}$. In addition, bulk bands form nodal points or nodal lines near $E_{\rm{F}}$ in momentum space (\textit{k} space)~\cite{burkov2011topological, potter2014quantum, fang2016topological, armitage2018weyl}. The nodal lines in topological semimetals can be gapped with inclusion of spin-orbit-coupling (SOC)~\cite{fang2015topological}. However, non-symmorphic symmetry can protect the nodal lines, keeping \rev{them} gapless even in the presence of SOC~\cite{zhao2016nonsymmorphic}. With this in mind, finding bulk superconductivity in topological semimetals with non-symmorphic space group symmetry is a promising path towards realizing topological superconductivity.

\begin{figure*}[t]
\centering
\includegraphics[width=16cm]{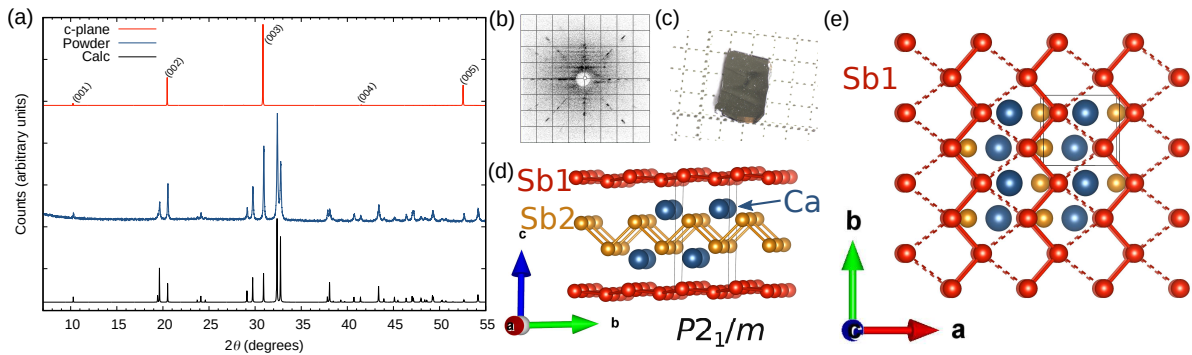}
\caption{(a) The X-ray diffraction pattern measured at room temperature of a \CS\ single crystal (top, red), powder (middle, blue), and the simulated powder pattern (bottom, black). (b) Laue pattern of a \CS\ single crystal. (c) Picture of a \CS\ single crystal on mm-size grid. (d) Crystal structure of \CS\ in the $bc$-plane showing Ca (blue) sites and the Sb zig-zag chains along the $b$-direction associated with the Sb1 (red) and Sb2 (orange) sites. (e) Crystal structure of \CS\ in the $ab$-plane showing zig-zag chains associated with the Sb1 site form a distorted square-net, where each of the distorted squares has two short (solid bonds) and two long (dashed bonds).}
\label{Fig1}
\end{figure*}

A number of materials with a square net of atoms in their crystal structure and non-symmorphic symmetry have been identified as topological semimetals in recent years, including SrMnSb$_2$~\cite{liu2017magnetic}, ZrSiS~\cite{schoop2016dirac} LaSbTe~\cite{singha2017magnetotransport, wang2021spectroscopic}, and GdSbTe~\cite{hosen2018discovery, lei2019charge, lei2021complex}. In these materials, a square net is formed such that the periodicity of the crystal structure is doubled compared to this square net itself. The doubling of the unit cell compared to the squares leads to folding of bands derived from $p_x$ and $p_y$ states of the square net atoms, and leads to a semimetallic state~\cite{klemenz2019topological}, where the square net atom can be Si or Sb. The square net $RE$SbTe ($RE$= rare earth) and the $M$Sb$_2$ ($M$= rare earth, alkaline earth) groups of materials are closely related, where the Sb square-net is distorted in the latter. We find the $M$Sb$_2$ interesting, in particular \CS , where the topologically non-trivial character of the bands persists even when the Sb square-nets are distorted. Recently, it was theoretically predicted that \CS\ is a nodal-line semimetal, and that this state is protected by the non-symmorphic symmetry of the crystal structure~\cite{funada2019spin}.
Experimental reports of a metal-to-insulator-like transition~\cite{funada2019spin} and superconductivity~\cite{ikeda2020superconductivity} in \CS\ have drawn interest to this material. Those experimental reports have led us to pursue the growth of single crystals, a critical step to making many of the measurements needed to explore the intrinsic properties of \CS .

In this paper, we confirm the existence of bulk superconductivity in high quality single crystals of \CS\ and take several steps in identifying the nature of this superconducting state. Measurements of the phase diagram in magnetic field give us the fundamental superconducting length scales in the material: the London penetration depth and the superconducting coherence length. Low temperature specific heat measurements suggest unconventional aspects of the superconducting pairing state. Furthermore, we present band structure calculations demonstrating that states associated with the distorted Sb square-net are dominant near the Fermi level and thus play a central role in the superconducting state and electrical properties. We observe de Haas–van Alphen (dHvA) oscillations consistent with our calculated Fermi surfaces. In electrical resistivity measurements under magnetic fields, we observe a rise in resistivity at low temperature, consistent with previous reports. The magnetoresistance at low temperature reaches high values without saturation and the concentration of hole and electron carriers is almost equivalent below 50~K. All of these results show that superconductivity in \CS\ occurs in the context of a well compensated semimetal.


\section{Methods}

Single crystals of CaSb$_2$ were grown from Sb self flux. Pieces of Ca (Alfa Aesar, 99.98\,\%) and Sb (Chempur, 99.999\,\%) with a molar ratio of 1:3.75 were loaded into an Al$_2$O$_3$ crucible and sealed in a quartz tube with a pressure of 0.3\,atm argon. The mixture was slowly heated to 700\,$^\circ$C and held at this temperature for 12\,h. Subsequently, the mixture was cooled to 670\,$^\circ$C at a rate of 10\,K/h, followed by slow cooling to 592\,$^\circ$C with 1\,K/h. At this temperature, the crystals were separated from the flux by centrifuging at 2000-3000 rpm for about 15 s. Shiny, plate-like crystals with dimensions up to 3 x 3 x 0.5\,mm$^3$ were obtained by this method. Crystals are stable in air for several weeks. Phase purity and orientation of the crystals were checked by X-ray diffraction (XRD) using a Bruker D8 with Cu K$\alpha_1$ radiation (1.54056~\AA ). The composition and homogeneity were confirmed by means of energy dispersive X-ray  spectroscopy (EDX) using a Philips XL30 scanning
electron microscope equipped with a Bruker Quantax 200 energy-dispersion X-ray microanalysis system and an XFlash 6010 SDD detector.

Electrical resistivity measurements were performed with conventional four- and five-probe geometries using a Quantum Design Physical Property Measurement System (PPMS) equipped with a $^3$He/$^4$He dilution refrigerator insert. For these measurements Pt wires (25 $\mu$m) were attached to the sample with silver epoxy.
Measurements of the magnetic susceptibility were done using a Magnetic Property Measurements System 3 (MPMS3) also from Quantum Design, equipped with a $^3$He insert.
Specific heat measurements were performed using a PPMS (Quantum Design).

Electronic structure calculations were performed within the framework of density functional theory (DFT) as implemented in the package Wien2k~\cite{blaha2001wien2k}. 
The generalized gradient approximation with the PBE parametrization~\cite{perdew1996generalized} was used. 
The basis set size was set to R$_{mt}$K$_{max}$=8.5 and the irreducible Brillouin zone (BZ) was sampled with 2445\,k~points. The previously reported monoclinic crystal structure ($P2_1/m$, S.G.:11, \textit{a} = 4.746~\AA , \textit{b} = 4.177~\AA , \textit{c} = 9.084~\AA , $\beta=106.3^{\circ}$) was used~\cite{deller1976darstellung}.

\section{Results and Discussion}

The \CS\ single crystals used in this study were characterized using X-ray diffraction (XRD) and energy dispersive X-ray  spectroscopy (EDX) to confirm the quality of the samples. The powder XRD pattern measured on our sample, shown in Fig.~\ref{Fig1}a in blue, is consistent with the previously reported monoclinic crystal structure ($P2_1/m$, S.G.:11)~\cite{deller1976darstellung}, but we refined the lattice parameters of our samples to \textit{a} = 4.7404~\AA , \textit{b} = 4.1796~\AA , \textit{c} = 9.0712~\AA, and $\beta$ = 106.3\,°. For comparison, we show in black in Fig.~\ref{Fig1}a the calculated diffraction pattern based on our refined structure. The Laue measurement in Fig.~\ref{Fig1}(b) shows sharp peaks confirming the high quality of the \CS\ samples. EDX mapping shows a homogeneous distribution of Ca and Sb on the sample surface (Sup.~Fig.~1 in  in Supplemental Materials~\cite{SupMat}). The $c$-axis diffraction pattern is shown in red in Fig.~\ref{Fig1}a and confirms that the direction perpendicular to the plate-like crystals (Fig.~\ref{Fig1}c) corresponds to the $c$ axis of the crystal structure (shown in Fig.~\ref{Fig1}d and e). The structure contains two zig-zag chains of antimony corresponding to two distinct crystallographic sites, Sb1 and Sb2 shown in red and orange, respectively, in Fig.~\ref{Fig1}d. The Ca atoms occupy one site, which sits between these two chains of Sb, as shown in blue in Fig.~\ref{Fig1}d. The zig-zag chains associated with the Sb1 site form a distorted square net (Fig.~\ref{Fig1}e), where squares become quadrilaterals with two short and two long distances of 2.915~\AA\ (solid bonds) and 3.459~\AA\ (dashed bonds), respectively. The zig-zag chains associated with Sb2 site have Sb-Sb distances of 2.939~\AA\ and the chains are separated by 4.740~\AA . The non-symmorphic space group contains a 2$_1$ screw axis along the $b$ direction, which is related to the symmetry of the Sb chains in the distorted square net.

\begin{figure*}[tbh]
\centering
\includegraphics[width=16cm]{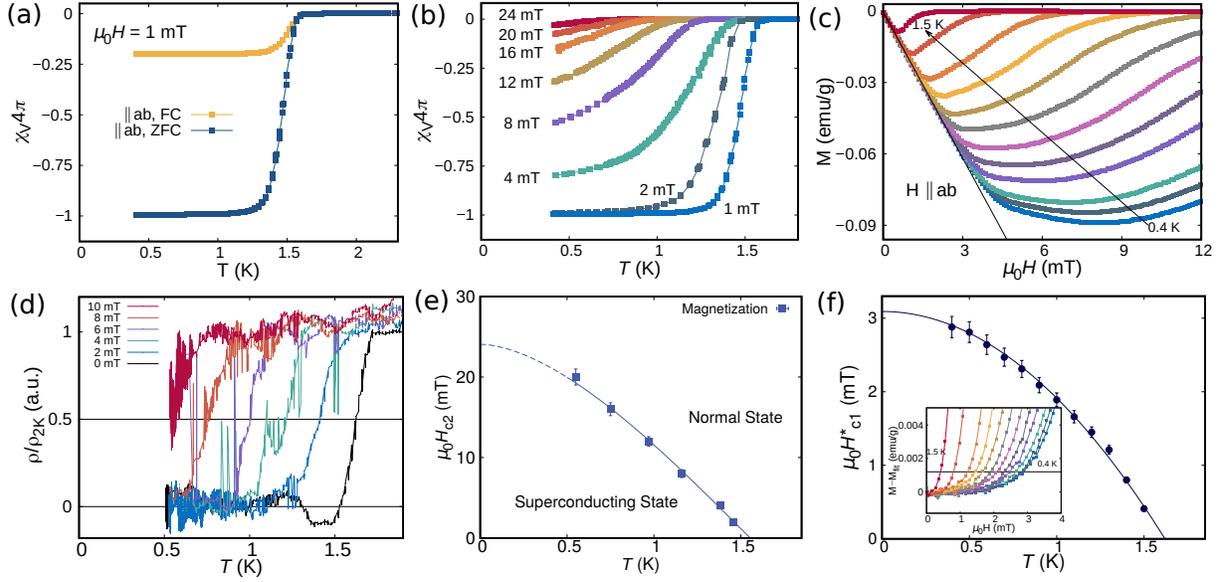}
\caption{(a) Temperature dependence of the dc magnetic susceptibility measured in a 1~mT field applied parallel to the $ab$-plane (along plane of crystal) with zero-field-cooling (ZFC) and field-cooling (FC) procedures, presented in terms of the shielding fraction. (b) Temperature dependence of the dc susceptibility measured with different applied fields using a zero-field-cooling procedure. (c) The magnetization ($M$) as a function of applied field measured at different temperatures below \Tc . A degaussing procedure was carried out between measurements, and a linear fit was applied to the low field region of the 0.4~K data, as discussed in the text. (d) Temperature dependence of longitudinal resistivity $\rho_\mathrm{xx}$ below 2~K measured in different magnetic fields, where the data are normalized by the 2~K data ($\rho_\mathrm{xx,2~\rm{K}}\sim1~\mu\Omega\rm{cm}$) measured in 0~mT. The superconducting transition is suppressed with increasing magnetic fields. (e) The \Tc\ values extracted from transitions in $\chi_V$ measured in different fields. Solid and dashed lines are WHH fits to the data used to estimate \mHct . (f) Lower critical field \mHco\ estimated using the magnetization data in (c) by subtracting the linear fit to 0.4~K data from all the curves, shown in inset. The cutoff is chosen as estimate of \mHco\ at each temperature.}
\label{Fig2}
\end{figure*}
\subsection{\label{SC}Superconductivity}

We characterize the superconducting state of \CS\ and show bulk superconductivity with sharp superconducting transitions, using magnetic susceptibility, specific heat, and resistivity measurements. The magnetic susceptibility at low temperature is shown in Fig.~\ref{Fig2}(a). The DC magnetization was measured between 0.4~K-1.9~K using zero-field-cooling (ZFC) processes and field-cooling processes (FC). The volume susceptibility ($\chi_V$) vs. temperature measured in a field of 1~mT is shown in Fig.~\ref{Fig2}(a), where $\chi_V$ was corrected for \rev{demagnetization effects due to sample shape using the expression for a slab-like geometry in Ref.~\cite{brandt1999irreversible}.} Both ZFC and FC measurements show evidence of bulk superconductivity with a strong diamagnetic signal that has an onset of 1.6~K in a field of 1~mT. The superconducting transition width $\Delta T$, defined as the temperature range over which the diamagnetic signal reaches 90\% of its maximum value below \Tc , is $\sim$0.25~K. \rev{The estimated superconducting volume fraction after demagnetization correction in the ZFC data is 100\%. We note that demagnetization calculated using a rectangular cuboid in Ref.~\cite{prozorov2018effective} resulted in a volume fraction exceeding 100\%. Thus, our data suggests near-perfect shielding in the superconducting state.} The difference between the ZFC and FC values of $\chi_V$ in our single crystal samples is suggestive of a type-II superconductor with flux pinning. However, the significant flux expulsion in the FC data suggests that pinning of flux vortices is weak. 

\begin{figure}[ht]
\centering
\includegraphics[width=8cm]{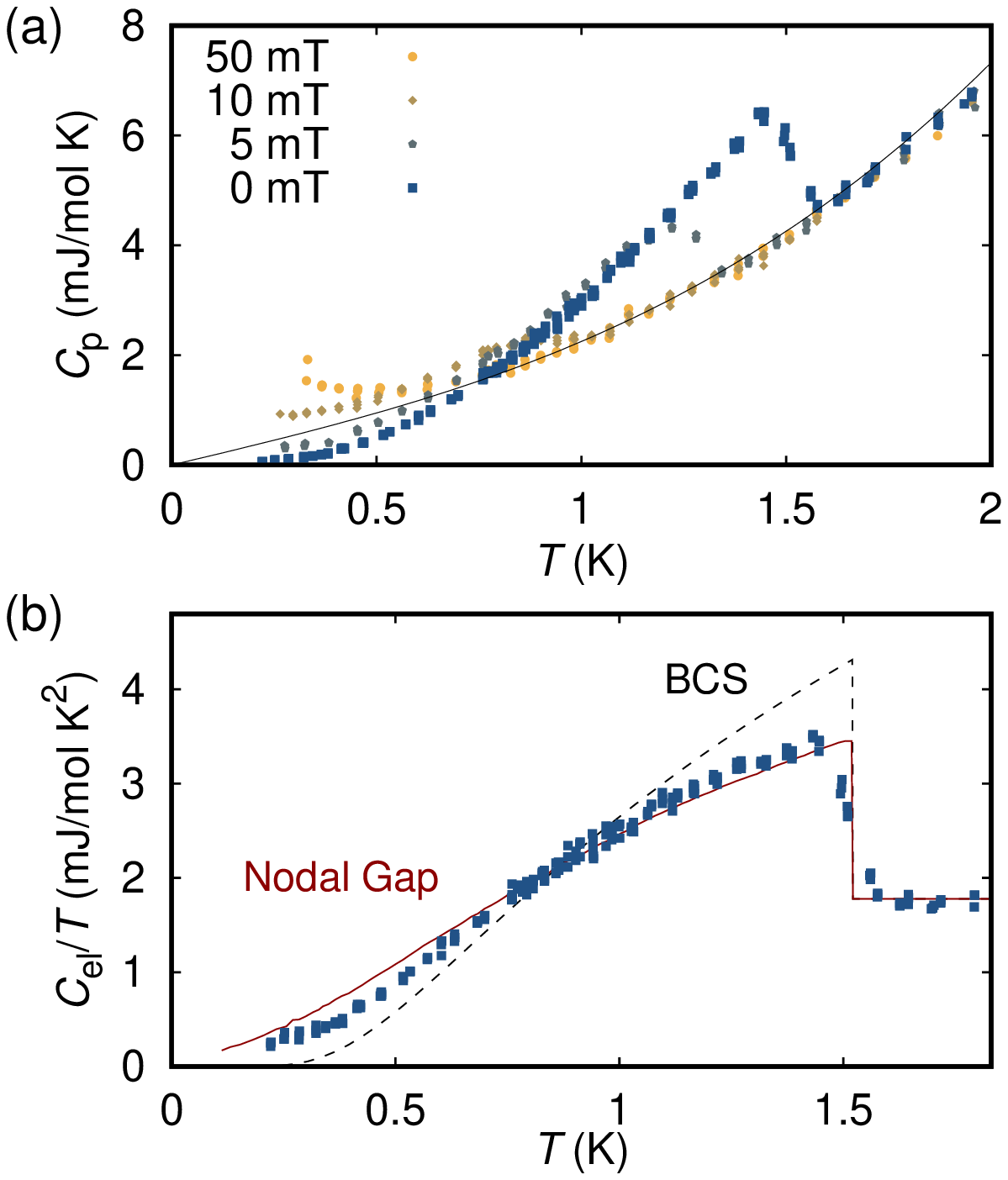}
\caption{(a) Temperature dependence of \Cp\ measured in varying magnetic fields. The superconducting transition seen at 1.52~K in 0~mT is suppressed with increasing magnetic field, emphasized in inset showing \Ce /$T$.  (b) Superconducting transition in the electronic part of the specific heat \Ce /$T$ as a function of $T$ in 0~T. Two fits are presented for a \Tc\ of 1.52~K: dashed black line is for weak-coupling BCS behavior, while solid red line is a theoretical curve for $d$-wave pairing on a simple cylindrical Fermi surface with line nodes~\cite{won1994d}.}
\label{Fig3}
\end{figure}

In order to map out the field-temperature phase diagram, we track the shift in \Tc\ for different applied magnetic fields in DC magnetization and specific heat measurements in Figs. \ref{Fig2}(b) and ~\ref{Fig3}(a), respectively. We measured a drop in electrical resistivity at the superconducting transition that is suppressed with applied field, as shown in Fig.~\ref{Fig2}(d), but due to noise in the data estimating \Tc\ at each applied field is difficult. Also, we measured the AC susceptibility of the \CS\ and the real ($\chi$'\textsubscript{AC}) and imaginary ($\chi$''\textsubscript{AC}) parts are shown in Sup.~Fig.~2 in Supplemental Materials~\cite{SupMat}. Although we measured a sharp superconducting transition with a peak in $\chi$''\textsubscript{AC} at 1.67~K, we find this measurement is sensitive to the onset of superconductivity rather than the bulk \Tc . From magnetization, \Tc\ is defined as the temperature where the diamagnetic signal reaches 5\% volume fraction at different magnetic fields. \rev{From specific heat, \Tc\ is defined as the mid point of the transition seen in \Ce , inset of Fig.~\ref{Fig3}(a).} The magnetic fields plotted against \Tc\ as extracted from both measurements are shown in Fig.~\ref{Fig2}(e). To estimate \mHct, we use the Werthamer-Helfand-Hohenberg (WHH) equation~\cite{werthamer1966temperature}.
We get \mHct\ of 21.5~mT from specific heat and 24.4~mT from magnetization. Using the \mHct\ of 24.4~mT, we estimate the coherence length $\xi_{\rm{GL}}$ using the equation

\begin{equation}
    \mu_0 H_{\text{c2}}(0) = \frac{\Phi_0}{2\pi\xi^2_{\text{GL}}}\
    \label{MuHPhi}
\end{equation}
\noindent
where $\Phi_0$ is the quantum flux $h/2e$. We get $\xi_{\rm{GL}}$ of 116~nm.

To measure \mHco\ we measured the field-dependent magnetization ($M$) at different temperatures below the critical temperature, as shown in Fig.~\ref{Fig2}(c). The low-field linear fit for the 0.4~K measurement is shown with a black line from the origin ($M_{\rm{fit}}$). This fit was subtracted from all the measured curves to construct a $M$-$M_{\rm{fit}}$ plot. The field where the magnetization deviated from linear response, taken as the black line in the inset of Fig.~\ref{Fig2}(f), is the uncorrected \mHsco\ for that temperature. All \mHsco\ are plotted against the corresponding temperature in Fig.~\ref{Fig2}(f) and are fitted to the equation

\begin{equation}
    \mu_0 H^*_{cl}(T) = \mu_0 H^*_{cl}(0) \left[1- \left(\frac{T}{T_c}\right)^2 \right]\
    \label{MuHcl}
\end{equation}

\noindent
where \mHsco\ is the lower critical field at 0~K and \Tc\ is the superconducting critical temperature. \mHsco\ was calculated to be 3.1~mT. After correcting for the demagnetization factor, \mHco\ was calculated to be 4.9~mT.

We estimate the penetration depth $\lambda$ using $\xi_{\rm{GL}}=116.1$~nm and \mHco\ $=4.9$~mT with the equation
\begin{equation}
   \mu_0 H_{\text{cl}} = \frac{\Phi_0}{4\pi\lambda^2} \left(\text{ln}\frac{\lambda}{\xi}+\frac{1}{2}\right)\
    \label{MuHclPhi}
\end{equation}
\noindent
we get $\lambda = 89.5$~nm for \CS . We calculate the Ginzburg-Landau constant $\kappa_{\rm{GL}}=\lambda/\xi_{\rm{GL}}=0.771\pm0.038$. Based on our calculated values, we classify \CS\ as a type-II superconductor since $\kappa_{\rm{GL}}> \frac{1}{\sqrt{2}}$.  However, it is just barely type-II, so this property will be sensitive to defects and carrier density in this semi-metallic material. 

In Fig.~\ref{Fig3}(b), the transition seen in \Ce\ with \rev{a \Tc\ of 1.52~K in zero field} and a $\Delta T$ of 0.15~K provides further evidence for the bulk nature of the superconductivity. Magnetic fields push superconducting transitions seen in specific heat \Cp\ to lower temperatures (Fig.~\ref{Fig3}(a)). The phonon contribution to \Cp\ was determined by fitting the data measured in a field of 50~mT, where no transition is observed, between 0.55~K-4.0~K to the expression 

\begin{equation}
    C_p = \gamma _nT + \beta T^3\
    \label{CpT}
\end{equation}

The result of this fit is $\gamma_n = 1.778$~mJ~mol$^{-1}$~K$^{-2}$ and $\beta = 0.471$~mJ~mol$^{-1}$~K$^{-4}$. The phonon contribution represented by the cubic term $\beta T^3$ was subtracted from the total \Cp\ to evaluate the electronic part of the specific heat \Ce .  The electronic contribution to specific heat \Ce\ divided by temperature is plotted as a function of temperature in Fig.~\ref{Fig3}(b). The flat part of the graph above 1.55~K corresponds to a constant $\gamma$ above the transition temperature consistent with metallic behavior. At \Tc\ a jump is observed corresponding to a second-order phase transition into the superconducting state, and the narrow transition further confirms the high-quality of the crystals. We observed the superconducting transition in $\chi_V$ and \Ce\ measurements on multiple crystals, grown in different batches, where the \Tc\ as defined above differs by less than 1\%.

The Debye model is used to calculate the Deybe temperature $\Theta_D$

\begin{equation}
    \Theta_D = \left(\frac{12\pi^4}{5\beta} nR\right)^{1/3}\
    \label{ThtaD}
\end{equation}
\noindent
where $n=3$ and $R$ is the gas constant 8.314~J~mol$^{-1}$~K$^{-1}$. The Debye temperature was calculated to be $\Theta_D=231~K$, and with $\Theta_D$ and \Tc , we calculate the superconducting parameter $\lambda_{ep}$ using the inverted McMillan~\cite{mcmillan1968transition} equation

\begin{equation}
    \lambda_{\text{ep}} = \frac{1.04+\mu^*ln \left( \frac{\Theta_D}{1.45T_c}\right)}{(1-0.62\mu^*) \text{ln} \left( \frac{\Theta_D}{1.45T_c}\right)-1.04}\
    \label{lamep}
\end{equation}
\noindent
where $\mu\rm{*}=0.10$ and \Tc~$= 1.60$~K. We find \rev{$\lambda_{ep}=0.458$ for \CS ,} suggesting it is a weak-moderate coupling superconductor. Using $\lambda_{ep}$, $\gamma$, and the Boltzmann constant $k_{\rm{B}}$, we calculate the electronic density of states at the Fermi energy $N(E_{\rm{F}})$ with the equation

\begin{equation}
    N(E_F)=\frac{3\gamma}{\pi^2 k_B^2(1+\lambda_{\text{ep}})}\
    \label{NEf}
\end{equation}
\noindent
$N(E_{\rm{F}})$  was estimated to be ~\rev{0.52 states eV$^{-1}$ per formula unit (fu) of \CS .}

\begin{figure*}[t]
\centering
\includegraphics[width=16cm]{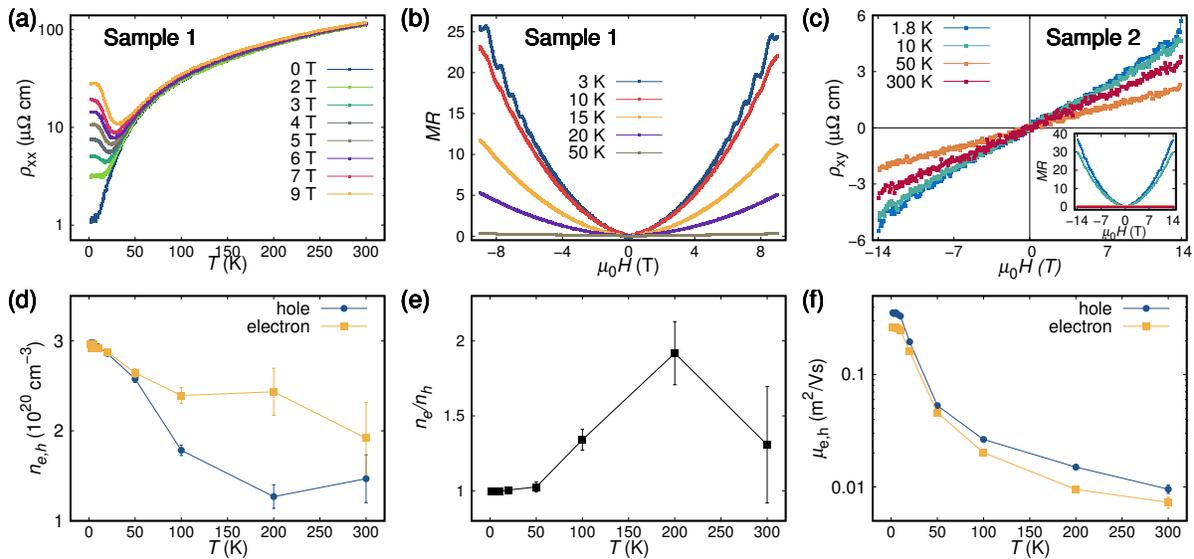}
\caption{Electrical transport data and analysis by two-band model. (a) Temperature dependence of the longitudinal resistivity $\rho_\mathrm{xx}$ for different magnetic fields with $\mathrm{I} \perp c$ in sample 1. (b) Magnetic field dependence of the resistance (MR) with $\mathrm{H}\parallel c$ in sample 1. (c) MR and Hall resistivity with $\mathrm{H}\parallel c$ in sample 2, respectively. (d-f) Charge carrier densities, their ratio, and carrier mobilities extracted from fits of MR and $\rho_\mathrm{xy}$ of sample 2 to the two-band model~\cite{pippard1989magnetoresistance}, respectively.}
\label{Fig4}
\end{figure*}

We calculate the mean free path ($l$) using the following equation
\begin{equation}
   l = 2.372 \times 10^{-14} \frac{\left(\frac{m^*}{m_e}\right)^2V_M^2}{N(E_F)^2 \rho}\
    \label{lmv}
\end{equation}
\noindent
\rev{where $N(E_{\rm{F}})=0.52$, the molar volume $V_{\rm{M}}=52.0~\rm{cm}^3/\rm{mol}$, $\rho=1~\mu\Omega\rm{cm}$ (Fig.~\ref{Fig4}(a)), and using $m^*/m_e=0.165$, based on dHvA oscillation analysis in Sec.~\ref{SM}, we get $l=65.3~\rm{nm}$. The resulting value of $\xi_{\rm{GL}}/l$ is 1.78, places \CS\ at the border between the clean limit and dirty limit.}


The magnitude of the specific heat jump $\Delta C/C_{en} \sim 0.95$, $C_{en}$ being the electronic specific heat in the normal state,  is about 34\% smaller than the value of 1.43 expected for a BCS model in the weak-coupling limit. In Fig.~\ref{Fig3}(b) we show the BCS expectation of a jump at \Tc\ and the exponential decay of \Ce\ below \Tc\ with a black dotted line, and it is clear that the data in \CS\ are not well explained with the BCS model. As we have near-perfect shielding from the Meissner signal in the magnetization measurements for multiple samples, we exclude the possibility of sample inhomogeneity as the origin of this smaller jump in the specific heat. \rev{We attempt a fit to the data based on the alpha model~\cite{padamsee1973quasiparticle,johnston2013elaboration}, and we find that experimental data deviates from this model for $\alpha = 1.425~\&~1.50$, as shown in Sup.~Fig.~3.} More importantly, the decay in \Ce\ $/T$ with decreasing temperature is weaker than that expected for a BCS superconductor. Even at the lowest temperature of 0.22~K the value of \Ce\ is not negligible and continues to be temperature-dependent. This unusual behaviour in \Ce\ suggests the presence of a small gap component, due perhaps to extreme gap anisotropy, differing gaps on different bands, or gaps with nodes. 

\begin{table}[t]
\centering
\caption{Superconducting parameters derived from our measurements of \CS.}
\label{SCTable}
\setlength{\extrarowheight}{8pt}
\begin{tabular}{|c c c|}
\hline
\textbf{Parameter} & \textbf{Units}      & \textbf{Value}  \\
\hline
\Tc        & K                   & 1.60    \\
\mHco    & mT                  & $4.9\pm0.4$   \\
\mHct    & mT                  & $24.4\pm0.7$  \\
$\xi_{\rm{GL}}$       & nm                  & 116   \\
$\lambda$        & nm                  & 89.5  \\
$\kappa_{\rm{GL}}$       & -                   & $0.771\pm0.038$   \\
$\gamma_n$        & mJ mol$^{-1}$ K$^{-2}$        & 1.778  \\
$\beta$         & mJ mol$^{-1}$ K$^{-4}$        & 0.471  \\
$\theta_D$        & K                   & 231.3  \\
$\lambda_{ep}$      & -                   & 0.458 \\
$N(E_{\rm{F}})$     & states eV$^{-1}$ per fu & 0.52 \\
\hline
\end{tabular}
\end{table}


We consider a model at the extreme end of possible gap anisotropy, one that includes line nodes in the superconducting gap~\cite{won1994d}, shown in red in Fig.~\ref{Fig3}(b). In this model, the average gap magnitude is smaller than that in the fully gapped case, and this is reflected in a reduced specific heat jump. Also, at low temperature the $T$ dependence of \Ce\ $/T$ changes from exponential to power law~\cite{maeno2011evaluation}, because the existence of nodes allows for the thermal excitation of quasiparticles down to very low temperatures. This nodal-gap function was recently suggested for a Bi$_2$Se$_3$ related material, namely Cu-intercalated (PbSe)$_5$(Bi$_2$Se$_3$)$_6$~\cite{sasaki2014superconductor}, considered to potentially be topological superconductors. In \CS\ this model fits better to the superconducting jump at \Tc , but the low temperature experimental data has lower values than expected in this model. The data clearly do not fit the BCS values or the simplest line-node model, which suggests strong anisotropy in the gap function. This may be related to one of the Fermi surfaces in our calculation \rev{being} dominated by Sb chains in the distorted square-net (Sb1 site), discussed in Sec.~\ref{SM}. We note the recent report of $^{121/123}$Sb-nuclear quadrupole resonance (NQR) measurement in a polycrystalline sample of \CS , which suggests $s$-wave superconductivity, in part based on observing a coherence peak below \Tc\ \cite{swavetakahashi}. The suggestion of exponentially-activated behaviour in \rev{those} NQR data at low temperatures is not clear, as is also the case for our specific heat data. A coherence peak was observed in nuclear magnetic resonance (NMR) measurements of the unconventional superconductors $\rm{CePt}_3\rm{Si}$~\cite{yogi2006evidence,yogi2006195pt} and $\rm{Li}_2(\rm{Pd}_{1-x}\rm{Pt}_x)_3\rm{B}$~\cite{harada201011b,shimamura2007nmr}, which is composition dependent in the \rev{latter}. Ultimately the resolution of the gap structure in \CS\ will require lower temperature measurements on single crystal samples, and a variety of spectroscopic probes.

\CS\ seems to fall at the border of type-I and type-II superconductivity. YbSb$_2$, which crystalizes in a higher symmetry crystal structure ($Cmcm$, S.G.:63), has been reported to be a type-I superconductor~\cite{zhao2012type}. The different crystal structures of YbSb$_2$ and \CS\ may be important to the nature of the emergent superconducting state in these compounds.
Studying the effect of pinning centers on flux-trapping in \CS\ and further improving the quality of grown crystals will help us clarify whether it is intrinsically a type-I or type-II superconductor, as well as further probing the presence of gap anisotropy. This will be the subject of future studies. A summary of observed superconducting parameters of \CS\ is presented in Table~\ref{SCTable}. \rev{We firmly  established and characterised bulk superconductivity in \CS . In the next section we place this superconductivity in the context of an unusual semimetal by examining the consistency of the calculated band structure with our transport measurements and de Haas–van Alphen (dHvA) oscillations.}

\begin{figure*}[t]
\centering
\includegraphics[width=16cm]{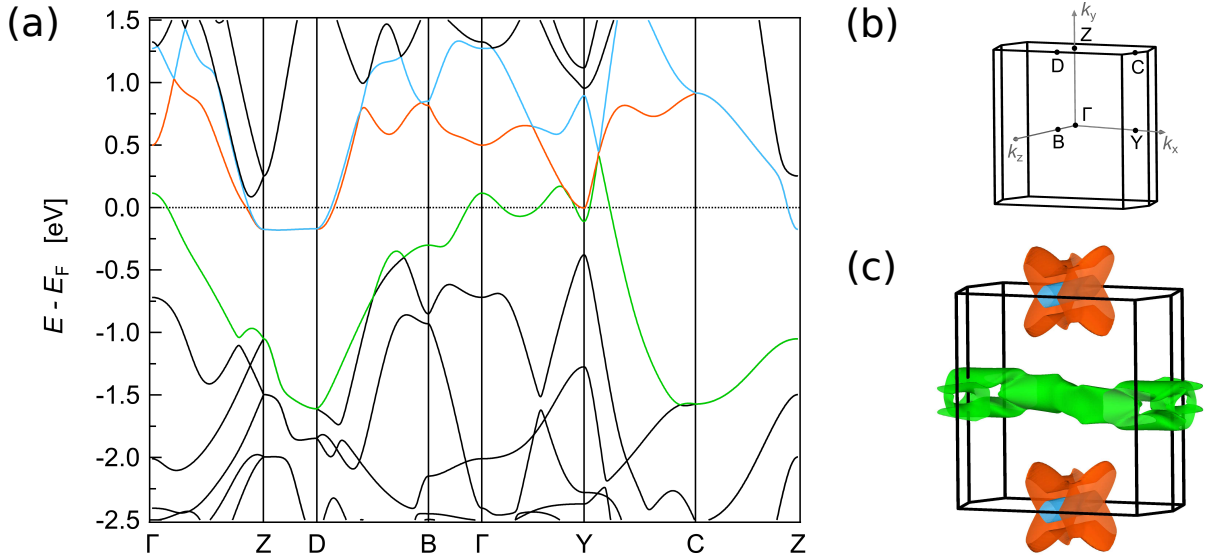}
\caption{Electronic structure of CaSb$_2$ from density functional theory calculations. (a) Band structure with bands crossing $E_\mathrm{F}$ colored, spin-orbit coupling (SOC) not included. (b) Brillouin zone of CaSb$_2$. (c) Three-dimensional visualization of the Fermi surface with SOC formed by one hole-like and two electron-like pockets. The color coding corresponds to that in panel (a).}
\label{Fig5}
\end{figure*}

\subsection{\label{SM}Semimetallic Properties and Band Structure}
The resistivity measurements reveal metallic behavior with a longitudinal resistivity $\rho_\mathrm{xx}$ of ca. 1\,$\mu\Omega\cdot$cm at 2\,K (Fig.~\ref{Fig4}a). The high residual resistivity ratio $\rho_\mathrm{300~\rm{K}}/\rho_\mathrm{2~\rm{K}}$ of ca. 100 indicates the good quality of the crystals. 
At low temperature, the resistivity rises rapidly with increasing field. While the zero field resistivity has the temperature dependence of a good metal, in higher fields the resistivity begins to increase with decreasing temperature as in an insulator. The rise in resistivity with decreasing temperature in an applied field larger than $3~\rm{T}$ plateaus for temperatures below $\sim10~\rm{K}$. This behavior in our measurements on single crystals of \CS\ is similar with those measurements reported on polycrystalline samples~\cite{fu2008superconducting}, as well as other topological semimetals such as WP$_2$~\cite{wang2017large}, NbAs$_2$~\cite{wang2016resistivity}, and ZrP$_2$~\cite{bannies2021extremely}. As shown in Fig.~\ref{Fig4}b, the magnetoresistance MR, which is defined as $MR=\rho_\mathrm{xx}(B)/\rho_\mathrm{xx}(B=0)-1$, remains unsaturated up to 9\,T, where it reaches large values of 2500\% at 2\,K. The field dependence of the MR can be described by a power law $MR=\mathrm{a}+\mathrm{b}\cdot B^\mathrm{c}$ with $\mathrm{c}=1.79$. This behavior is reproduced in a second sample (see Fig.~\ref{Fig4}c inset). The Shubnikov-de Haas oscillations superimposed on the MR background (Fig.~\ref{Fig4}b) further demonstrate good sample quality but are not analyzed here due to the low number of oscillations observed. Instead we will present de Haas-van Alphen (dHvA) oscillations in detail below.
Together, the field-driven metal-to-insulator-like transition and the large and unsaturated MR suggest a semimetallic state in CaSb$_2$.

To verify this scenario, we measured the Hall resistivity $\rho_\mathrm{xy}$ in the same field and temperature ranges as the MR~(Fig.~\ref{Fig4}c). \rev{While $\rho_\mathrm{xy}(B)$ is roughly linear for all temperatures, it cannot be fit with a semi-classical one-band model because $\rho_\mathrm{xx}$ is independent of the field in this model, which is contrast to our data (see Fig.~\ref{Fig4}b). Instead,} the data can be fitted with the semi-classical two-band model~\cite{pippard1989magnetoresistance}: 

\begin{equation}
    \rho_\mathrm{xx}=\frac{1}{e}\frac{(n_\mathrm{h}\mu_\mathrm{h}+n_\mathrm{e}\mu_\mathrm{e})+(n_\mathrm{h}\mu_\mathrm{e}+n_\mathrm{e}\mu_\mathrm{h})\mu_\mathrm{h}\mu_\mathrm{e}B^2}{(n_\mathrm{h}\mu_\mathrm{h}+n_\mathrm{e}\mu_\mathrm{e})^2+(n_\mathrm{h}-n_\mathrm{e})^2\mu_\mathrm{h}^2\mu_\mathrm{e}^2B^2}
\end{equation}
and
\begin{equation}
    \rho_\mathrm{xy}=\frac{B}{e}\frac{n_\mathrm{h}\mu_\mathrm{h}^2-n_\mathrm{e}\mu_\mathrm{e}^2+(n_\mathrm{h}-n_\mathrm{e})\mu_\mathrm{h}^2\mu_\mathrm{e}^2B^2}{(n_\mathrm{h}\mu_\mathrm{h}+n_\mathrm{e}\mu_\mathrm{e})^2+(n_\mathrm{h}-n_\mathrm{e})^2\mu_\mathrm{h}^2\mu_\mathrm{e}^2B^2}
\end{equation}
based on the carrier densities $n_\mathrm{e,h}$ and mobilities $\mu_\mathrm{e,h}$ for electrons and holes. By simultaneously fitting $\rho_\mathrm{xx}$ and $\rho_\mathrm{xy}$ to the two-band model \rev{(see Sup.~Fig.~4 for an exemplary fit)}, we extracted the carrier densities and mobilities of the charge carriers, which are shown as functions of temperature in Fig.~\ref{Fig4}d-f (the full dataset is shown in \rev{Sup.~Fig.~5} in Supplemental Materials~\cite{SupMat}). While electrons are the majority charge carriers at high temperatures, the carrier densities are essentially compensated below 50\,K with $n_\mathrm{e}/n_\mathrm{h}=0.99(1)$ at 2\,K. The carrier mobilities increase monotonically with decreasing temperature, presumably due to reduced scattering of charge carriers by phonons. The high values of approximately 3000\,cm$^2$/Vs at 2\,K further indicate good crystal quality. Based on the two-band model fits, we conclude that CaSb$_2$ is a compensated semimetal in which large MR arises at low temperature from the combination of charge carrier compensation and high carrier mobility.

\begin{figure*}[t]
\centering
\includegraphics[width=16cm]{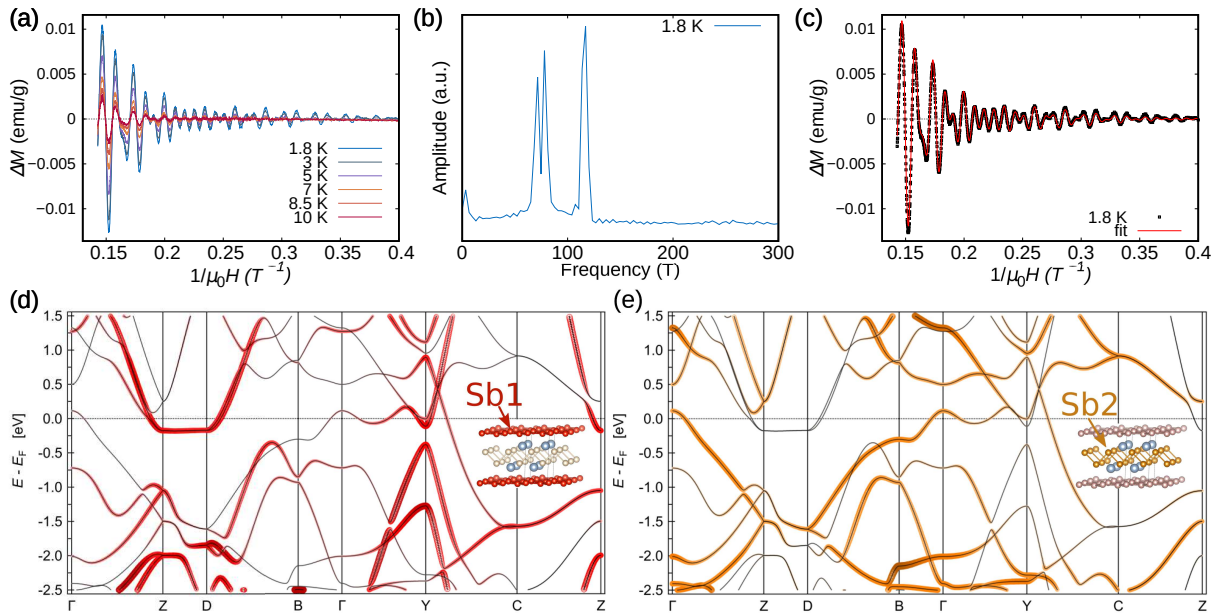}
\caption{Quantum oscillations and fat band representations based on density functional theory calculations. (a) Oscillatory component $\Delta M$ corresponding to the de Haas-van Alphen (dHvA) oscillations at different temperatures. (b) Fast Fourier Transform spectrum of the dHvA oscillations at 1.8\,K. (c) Fit of eq.~\ref{eqLK} to the dHvA oscillations at a representative temperature of 1.8\,K, using three fundamental frequencies. (d, e) Fat band representations of the band structure of CaSb$_2$. The size of symbols represents the projected weight of the Sb1 and Sb2 atomic wave functions, respectively.}
\label{Fig6}
\end{figure*}

We confirmed our experimental findings by band structure calculations based on density functional theory (DFT). The resulting band structure of CaSb$_2$ is shown in Fig.~\ref{Fig5}(a). The Fermi surface is composed of one hole-like pocket centered in the $k_\mathrm{y}=0$ plane and two electron-like pockets at the edge of the Brillouin zone (BZ), see Fig.~\ref{Fig5}c. While the hole pocket exhibits a complex three-dimensional shape, the two electron-like bands have weak dispersion along $k_\mathrm{z}$, giving rise to a quasi-two-dimensional character. 
The presence of two electron pockets in the calculations seemingly contradicts our analysis of the magnetotransport. However, the volume of the smaller electron pocket is only 1/8 of the volume of the bigger electron pocket. We argue that the two electrons pockets can be effectively described as one combined electron pocket if they have similar average carrier mobilities. This assumption is validated by the good quality of the two-band model fits to the experimental data. 
Attempts to fit the data with a three-band model containing two electron pockets and one hole pocket did not improve the quality of the fit and were therefore discarded.




\begin{table*}
\centering
\caption{Extracted parameters from the fit of the de Haas-van Alphen oscillations with eq.~\ref{eqLK}. Extremal areas $A$ were calculated based on frequencies $F$ by use of the Onsager relation~\cite{shoenberg2009magnetic}. Effective masses $m^*$ and Dingle temperatures $T_\mathrm{D}$ were used to calculate quantum relaxation times $\tau_\mathrm{q}=\hbar/(2\pi k_\mathrm{B}T_\mathrm{D})$ and quantum mobilities $\mu_\mathrm{q}=e\tau_\mathrm{q}/m^*$.}
\label{dHvATable}
\setlength{\extrarowheight}{8pt}
\begin{tabular}{|cccccc|}
\hline
$F$ {[}T{]} & $A$ {[}10$^{-3}$\AA$^{-2}${]} & $m^*$ {[}$m_e${]} & $T_\mathrm{D}$ {[}K{]} & $\tau_\mathrm{q}$ {[}10$^{-13}$ s{]} & $\mu_\mathrm{q}$ {[}m$^2$/Vs{]} \\
\hline
71.5(1)   & 6.829(9)         & 0.168(1)      & 6.5(2)       & 1.88(7)              & 0.197(9)          \\
77.5(1)   & 7.398(7)         & 0.160(1)      & 6.3(2)       & 1.94(7)              & 0.213(9)          \\
115.8(1)  & 11.050(3)        & 0.166(1)      & 5.2(1)       & 2.32(4)              & 0.247(6)          \\
\hline
\end{tabular}
\end{table*}

To complement our findings about the Fermi surface based on electrical transport experiments, we have studied the magnetization in the normal state at low temperatures. At 1.8\,K, we observe de Haas-van Alphen (dHvA) oscillations in fields larger than 2.5\,T for $B\parallel c$. In Fig.~\ref{Fig6}(a), we show the oscillatory component of the dHvA oscillations as a function of inverse magnetic field at selected temperatures between 1.8\,K and 10\,K. The fast Fourier transform (FFT) spectrum of the data at 1.8\,K, shown in Fig.~\ref{Fig6}b, reveals three peaks at $F_\alpha=(70\pm4)$\,T, $F_\beta=(78\pm4)$\,T and $F_\gamma=(116\pm4)$\,T.
Given the limited number of periods collected, and the presence of closely-overlapping frequencies, it is advantageous to fit the dHvA oscillations to a standard model for a Fermi liquid in order to avoid limitations of the FFT method~\cite{ramshaw2011angle}. In this procedure, we fitted the oscillatory component $\Delta M$ of the dHvA oscillations as a function of magnetic field and temperature to the Lifshitz-Kosevich formula~\cite{shoenberg2009magnetic}:

\begin{equation}
    \Delta M\sim-B^{1/2}R_\mathrm{T}R_\mathrm{D}R_\mathrm{S}\sin{\left(2\pi\left(\frac{F}{B}-\frac{1}{2}+\phi\right)\right)}
    \label{eqLK}
\end{equation}
with $R_\mathrm{T}=\frac{XT}{\sinh(XT)}$, $R_\mathrm{D}=\exp{(-XT_\mathrm{D})}$, and $R_\mathrm{S}=\cos{\left(\pi g \frac{m^*}{2m_0}\right)}$, where $X=2\pi^2k_\mathrm{B}m^*/(\hbar eB)$. $T_\mathrm{D}$ is the Dingle temperature and $\phi$ is a phase factor that depends on the dimensionality of the Fermi surface and the Berry phase among other factors.

Accordingly, the temperature and magnetic field dependent damping of the oscillations are described by $R_\mathrm{T}$ and $R_\mathrm{D}$. Fitting $\Delta M(B,T)$ to eq.~\ref{eqLK} therefore gives access to the effective masses and Dingle temperatures. Furthermore, it allows us to refine the initially imprecise frequencies from the FFT. In Fig.~\ref{Fig6}c, we show a representative part of the fit at a constant temperature of 1.8\,K, which demonstrates the good quality of the fit. The data extracted from the fit at all temperatures are summarized in Tab.~\ref{dHvATable}. 

The three fundamental frequencies are 71.5(1)\,T, 77.5(1)\,T, and 115.8(1)\,T. Using the Onsager relation $A_\mathrm{F}=\frac{2\pi e}{\hbar}F$~\cite{shoenberg2009magnetic}, we find that the extremal orbits associated with these frequencies cover 0.357(5)\%, 0.387(4)\%, and 0.578(2)\% of the BZ projected along [001], respectively. The small values suggest small volumes of these Fermi surface pockets, in line with the semimetallic character. All three orbits show similar low effective masses of \textasciitilde\,0.16\,$m_e$ and the quantum mobilities range between \textasciitilde\,2000 and 2400\,cm$^2$/Vs, indicative of high sample quality. Altogether, the small extremal areas, low effective masses and high quantum mobilities are consistent with the semimetallic character implied by the transport measurements.

Next, we compare the extremal orbits from dHvA oscillations with those predicted by DFT calculations. We do not find a quantitative match between theory and experiment. The mismatch can be caused by two possibly concomitant reasons: (i) the as-grown single crystals are self-doped, resulting in an experimental Fermi energy different from the theoretical one or (ii) the functional used in DFT calculations does not describe the semimetallicity accurately, resulting in a wrong Fermi surface.  Without studies of the angle-dependence of the dHvA oscillations, it is impossible to deduce the reason for the mismatch. Future studies should therefore aim at obtaining a 3D Fermi surface from quantum oscillations. Importantly, our experimental results confirm the semimetallicity independent of the DFT calculations and thus remain valid despite the mismatch with the DFT calculations.

The bands around the Fermi level are dominated by contributions from Sb states, as shown by the fat band representations of Sb1 and Sb2 atoms in Fig.~\ref{Fig6}(d) and (e), respectively. The two electron bands crossing the Fermi level along the $\Gamma$-Z line close to the Z point have strong Sb1 character, whereas the hole band centered around the $\Gamma$ point is dominated by Sb2 character. The Ca contribution to these bands is very small as shown in \rev{Sup.~Fig.~6} in Supplemental Materials~\cite{SupMat}. The electron bands crossing the Fermi level along C-Z form nodal lines in the $k_y=\pi$ plane in the presence of SOC (see \rev{Sup.~Fig.~7} in Supplemental Materials~\cite{SupMat}), in agreement with previous calculations~\cite{funada2019spin}. These electron bands have a strong Sb1 character near the Fermi level and significant Sb2 character at higher energy; the Sb1 site forms the distorted square-net in \CS . The similarities in the crystal structure between the $M$Sb$_2$ distorted square net family and the $M$SbTe square net family may result in similarities in the band structure of these materials. We highlight that the periodicity of atoms compared with the periodicity of the unit cell in a particular direction results in non-symmorphic symmetry, which is important for the nodal-line semimetal state in both cases, rather than the perfect square-net of Sb in $M$SbTe. The connection between the nodal-line semimetal states in both classes of materials warrants further theoretical and experimental investigation. The combined evidence for semimetallic character from electrical transport and dHvA oscillations support our calculated band structure, which includes non-trivial band topology along the C-Z direction at the $k_y=\pi$ plane. 

Interestingly, \CS\ is one of only a handful of materials that displays at ambient pressure both a large MR (i.e. semimetallic character) and superconductivity, see Tab.~\ref{MRSCTable}. These materials include MoTe$_2$~\cite{qi2016superconductivity}, TaSe$_3$~\cite{sambongi1977superconductivity,saleheen2020evidence}, and $\alpha$-Ga~\cite{chen2018large}. Topological states were experimentally observed by angle-resolved photoemission spectroscopy (ARPES) in MoTe$_2$~\cite{deng2016experimental} and TaSe$_3$~\cite{lin2021visualization}.
Within superconducting semimetals, \CS\ offers a good balance between a moderately high \Tc\ of 1.6~K and large MR of 25 at 2~K and 9~T, similar to TaSe$_3$. In contrast, the low \Tc\ of MoTe$_2$ is difficult to reach, posing a significant disadvantage for applications. Among the materials listed in Tab.~\ref{MRSCTable}, $\alpha$-Ga has superior characteristics both in terms of \Tc\ and MR, but Ga is difficult to handle due to its low melting point of 29.8~$^{\circ}$C~\cite{sostman1977melting}. The combination of \Tc , MR, and stability of \CS\ make it an excellent candidate for exploring the interplay between superconductivity and semimetallicity.

\begin{table}[tbhp]
\centering
\caption{Materials that exhibit large magnetoresistance (MR) and superconductivity at ambient pressure, their superconducting transition temperatures \Tc , and MR values at 9\,T.}
\label{MRSCTable}
\setlength{\extrarowheight}{8pt}
\begin{tabular}{|cccc|}
\hline
Material & \Tc\ [K] & MR & Ref. \\
\hline
MoTe$_2$ & 0.1 & $\sim40$ at 1.4\,K, 9\,T & \cite{qi2016superconductivity}\\
$\alpha$-Ga & 0.9 & $\sim1.6\cdot$10$^4$ at 2\,K, 9\,T & \cite{chen2018large} \\
TaSe$_3$  & $\sim2.1$ & $\sim30$ at 1.9\,K, 9\,T & \cite{sambongi1977superconductivity,saleheen2020evidence}\\
CaSb$_2$  & 1.6 & 25 at 1.8\,K, 9\,T & This work \\
\hline
\end{tabular}
\end{table}

In theoretical work, it was suggested that the nodal-line structure of the normal state Fermi surface of \CS\ could survive in the superconducting state when weak-coupling is present~\cite{ono2021z}. This allows for realizing topological superconductivity in \CS . We find experimentally that superconductivity in single crystals of \CS\ is in the weak-moderate coupling limit with $\lambda_{ep}=0.456$ and the decay in the electronic part of the specific heat \Ce\ in the superconducting state down to 0.22~K is slower and does not follow the exponential decay expected for a conventional BCS superconductor. Future experiments will help us clarify the extent of influence of topology in the normal state of \CS\ on the superconducting properties.



\section{Conclusions}

\rev{In this Article, which is the first report on single crystals of \CS , we report evidence for bulk superconductivity in \CS\ with an onset of about 1.6~K.} The \Cp\ jump at \Tc\ is inconsistent with that expected in a BCS superconductor, and the decay in the \Ce\ at low temperature hints at anisotropy of the superconducting gap function. Our transport measurements support a well-compensated semimetal picture for \CS , where electrons and holes are almost perfectly compensated below 20 K. The high quality of the single crystals allows us to measure de Haas-van Alphen (dHvA) oscillations and we identify three frequencies associated with a small Fermi surface. Measuring the intrinsic superconducting and compensated semimetal properties of \CS\ was only possible due to availability of single crystals. Our calculations reveal that bands at the Fermi level are dominated by contributions from the distorted Sb square-net. The exact nature of the topologically non-trivial band crossing can be elucidated in future experiments, including ARPES. 
The family of square-net materials and diantimonides with distorted square-nets host a rich variety of properties, and the present work will serve as an important step towards connecting these two sub-fields. Our results promote \CS , and the family of $M$Sb$_2$, for further investigation of topological properties and superconductivity. 

\rev{\textit{Note added.} After submitting this manuscript, we noticed a report on the properties of polycrystalline \CS\ showing a peak in \Tc\ under pressure, and suggesting an unconventional nature of the superconductivity~\cite{kitagawa2021peak}.}

\section{Acknowledgements}
We thank Mengxing (Ketty) Na for help with the Laue measurement, Jacob Kabel for help with the EDX measurement, Silvia Luescher-Folk for help with low temperature electrical transport measurement, and Ilya Elfimov for discussion on DFT calculations. MO acknowledges the support by Stewart Blusson Quantum Matter Institute and the Max Planck-UBC-UTokyo Center for Quantum Materials. JB, DAB, and MCA acknowledge the support by the Natural Sciences and Engineering Research Council of Canada (NSERC).

\clearpage
\renewcommand{\thesection}{S\Roman{section}}
\setcounter{section}{0}
\renewcommand{\theequation}{S\arabic{equation}}
\setcounter{equation}{0}
\renewcommand{\thefigure}{S\arabic{figure}}
\setcounter{figure}{0}
\renewcommand{\thetable}{S\arabic{table}}
\setcounter{table}{0}

\bibliography{CaSb2}

\end{document}